# The dichotomy between 'practical' and 'theoretical' astronomy in ancient and late antique literature

Elio Antonello
*INAF- Astronomical Observatory of Brera*
*Italian Society for Archaeoastronomy*
elio.antonello@brera.inaf.it

**Abstract.** In Plato's dialogues *Republic* and *Laws*, the most important disciplines for the best education of the rulers of the city are identified with arithmetic, geometry and astronomy. Those disciplines, however, are not intended for practical applications, but to reach the truth and see the form of good. In particular, another dialogue, *Epinomis*, stresses on the relevance of astronomy itself as main discipline, since it coincides with the study of the gods, that is, the planets and the heaven. According to *Epinomis*, the wise astronomer does not observe the risings and settings of stars for practical applications such as the farmer calendar, but he studies the orbits of the planets. Therefore, the 'practical' astronomy of farmers appears intrinsically less important than the 'theoretical' astronomy, i.e. the study of the planetary motions. We discuss the possibly far-reaching negative implications of such a specific discrimination. We report some examples taken from Greek and Latin literature illustrating the difficulties of a coherent description of the risings and settings of stars that had been attempted by scholars, and probably had been of little help for farming. We conclude by pointing out the practical importance of astronomy (intended in a broad sense) even today, and of the dangers of the separation or discrimination of disciplines and sub-disciplines for the culture during the present global economic crisis.

**1. Introduction**

Among the ancient philosophers, Plato (428-348 BC) is generally recognized as one of the founders of Western philosophy. In the Dialogues he wrote about logic, ethics, religion, mathematics. In particular he discussed the ideal state or government, and the best education for its rulers. The most important disciplines were arithmetic, geometry and astronomy, since they studied the nature of the numbers and the true motions of the things in the sky, grasped by reason and thought, not by sight. Such a study should have been intended not for a practical application, but to see the truth and the form of good. In the Dialogue *Epinomis* there is a clear distinction or discrimination between the astronomy of the farmers (i.e. the farming calendar) and the study of planet motions (the planets are gods); we will call the former 'practical' astronomy, and the latter 'theoretical' astronomy. We will attempt to show the negative implications of that distinction, taking into account some examples taken from ancient and late antique literature. We will mention the problems of astronomical calendars used for farming, and some of their inconsistencies, as they appear in the works of Latin and Greek scholars. Finally, we will point out the dangers of the present distinction or separation of disciplines and sub-disciplines for the culture (e.g. Antonello, 2013b).

    We remark that the present study is not an essay on the history of philosophy, and it is not a critical discussion of the farming calendars and the related astronomical quandaries; it is just an attempt to point out some problems of the world of culture.

## 2. Plato's dialogues

In the following subsections we will quote at length the *Republic*, *Timaeus* and *Laws*, and then *Epinomis*. We think it is worth mentioning specific details of those dialogues, since they are of relevance for assessing the issue.

### *2.1 Republic*
In the ideal city depicted by Plato, the truly just society needs philosophic rulers, and the education of such philosophers is discussed in book VII of *Republic*. It begins with the allegory of the cave (corresponding to our visible world), where men live as prisoners, and the shadows they see are taken as the reality. The goal of the education is to free the prisoners from their false opinions and convictions which are based on such 'reality'. The philosophers have to make the ascent (outside the cave), and see the good, which is beyond the perceived reality (the one inside the cave). Once they have seen the good, they must conclude that it is the cause of all that is correct and beautiful in anything, that it produces both light and its source in the visible realm, and that in the intelligible realm it controls and provides truth and understanding. After that, the philosophers have to come back and live with the prisoners, for the benefit of the community.

Education isn't putting knowledge into souls that lack it, like putting sight into blind eyes (*Republic,* VII, 518b, c). The power to learn is present in everyone's soul, and education takes for granted that 'sight' is there, but it isn't turned the right way, so the education tries to redirect it appropriately (*Republic*, VII, 518d). What are the disciplines, apart from the dialectics, that draw the soul from the realm of becoming to the realm of what is? They have to do with the number, and not just in the practical way that is useful for a warrior ruler, but in such a way that a philosopher-ruler learns to rise up out of becoming and grasp being. The disciplines are therefore computation and arithmetic, plane and solid geometry, and finally astronomy.

Astronomy should not consist in a better awareness of the seasons, months, and years, which is no less appropriate for a general than for a farmer or a navigator (*Republic*, VII, 527d). It is not true that astronomy by itself compels the soul to look upward and leads it from things here to things there. According to Plato, indeed, astronomy as practiced by scholars makes the soul look very much downward; it is useful for practical applications, but a philosopher shall not be afraid to propose things that apparently are of no utility to most people. Plato cannot conceive of any subject making the soul looking upward except one concerned with that which is, and that which is invisible; the truth consists just in that invisible thing (*Republic*, VII, 529b).

### *2.2 Timaeus*
In the *Timaeus*, Plato presents an account of the formation of the universe. The god created firstly its soul, and then he made a moving image of eternity according to number. Time came to be after the model of that which is eternal. The good craftsman brought into being the Sun, the Moon and five other stars, for the begetting of time. He made the fixed stars and spread these gods throughout the whole heaven. Then there were two other generations of gods, "who are called by names we know", beginning with Earth and Heaven (*Timaeus*, 41a). Their task was to make the mortal creatures, weaving what is mortal to the immortal seed (the soul) sowed by the craftsman. The eyes were the first of the organs of the head to be fashioned by them (*Timaeus*, 45b). Our sight has proved to be a source of supreme benefit to us, in that none of our statements about the universe could ever have been made if we had never seen the heaven. Our ability to see the periods of day and night, of months and of years, of equinoxes and solstices, has led to the invention of number, and has given us the idea of time and opened the path to inquiry into the nature of the universe. These pursuits have given us philosophy, a gift from the gods to the mortal race whose value neither has been nor ever will be surpassed. The god "invented sight and gave it to us so that we might observe the orbits of intelligence in the universe and apply them to the revolutions of our own understanding. For there is a kinship between them, even though our revolutions are disturbed, whereas the universal orbits are undisturbed" (*Timaeus*, 47a - c).

## 2.3 Laws

In the *Laws*, the old philosopher discusses the codes of law of a new city, and the book VII deals with the education. Three disciplines are mentioned, computation and study of numbers, measurements of lines, surfaces and solids, and the mutual relationship of the heavenly bodies as they revolve in their courses. These subjects must be studied in minute detail not by the general public, but only by a chosen few, even though those disciplines would be actually important for all (*Laws,* VII, 817e-818a).

Plato affirms that he is not just deterred by the way people commonly neglect that subject, but he is even more appalled at those who have actually undertaken those studies, but in the wrong manner. Indeed, total ignorance over an entire field is never disastrous; much more damage is done when a subject is known in detail, but has been improperly taught (*Laws*, VII, 819a). Plato appears particularly amazed at the general ignorance about commensurables and incommensurables, blushing not only for himself but also for the Greeks in general (*Laws,* VII, 819d-820b). There is, moreover, another intolerable thing concerning astronomy: people generally believe that the heavenly bodies never follow the same path; this is the reason for the name 'planets', i.e. wanderers. According to Plato, that is blasphemy: actually the great gods (Sun, Moon, planets) perpetually describe just one fixed orbit, although it is true that to all the appearances its path is always changing (*Laws*, VII, 821a-822a).

The book XII of the *Laws* describes the Nocturnal Council, its members (such as the ten Guardians of Laws), and their duties; the Council has to meet daily from dawn, to muse on legislation. The Guardians must have an advanced education. One of the finest fields of knowledge is theology, which includes the study of the motion of the heavenly bodies and the other objects under the control of reason, which is responsible for the order in the universe. No one who has contemplated all this with a careful and expert eye has in fact ever degenerated into such ungodliness as to reach the position that most people would expect him to reach. They suppose that if a man goes in for such things as astronomy and the essential associated disciplines, and sees events apparently happening by necessity rather than because they are directed by the intention of a benevolent will, he will turn into an atheist (*Laws*, XII, 966e-967a). Plato affirms that the situation is different from the times when thinkers regarded the heavenly bodies as inanimate. Even then, those who studied them really closely had an inkling that the remarkably accurate predictions about their behaviour would never have been possible if they were inanimate, and therefore irrational (*Laws*, XII, 967a,b).

According to Plato, a mortal can attain a truly religious outlook if he grasps the doctrine that reason is the supreme power among the heavenly bodies; that shall be indeed the case of the Authorities of the Nocturnal Council.

## 2.4. Epinomis

As suggested by the Greek name, this dialogue is an addition to the *Laws*. Generally, scholars doubt its authenticity, and it may be possible that a Plato's disciple, Philip of Opus, had written it. Cooper (1997, p. 1617), noting some discrepancies with the *Laws*, writes that, if such was the case, it would present one of the first 'Platonisms', carrying forward the 'spirit' of Plato's work while giving distorting emphases to various elements within it.

The goal of the dialogue is to give an answer to the question: what must a mortal learn in order to be wise? First, one has to go through all the other sciences that do not make those who possess them wise, and then identify the needed ones and learn them (*Epinomis,* 974d). The production of food, in combination with the knowledge of how to use it, is a noble pursuit, but it "will never succeed in making anyone completely wise, since this very thing – labelling production as wisdom – would lead to disgust at the products themselves. Nor will the cultivation of the entire earth make anyone completely wise: it is clearly not by art but by a natural capacity we have from God that we have all put our hands to working the earth" (*Epinomis,* 975b, c). Similarly, all the arts, crafts and technical abilities enable us to possess the necessities of life, but none of them makes anyone wise. If we compare a discipline with another, we will see that the one that has given the gift of number would have this effect. It is God himself that saves us by making this gift, and such a God is the heaven. By "decorating himself and making the stars revolve in himself through all their orbits, he brings about the seasons and provides nourishment for all. Together with the entirety of number, he also furnishes […] everything else that involves intelligence and everything that is good" (*Epinomis*, 977b). If the human race were deprived of number, we would never come to be intelligent in anything, and would be unable to give a rational account.

How did we learn to count? The first thing God caused to dwell in us was the capability to understand what we are shown. And of the things he shows us, what can we behold more beautiful than the day? Later, when we come to see the night, everything appears different. Since heaven never stops making these bodies ply their course night after night and day after day, he never stops teaching humans one and two, until even the slowest person learns well enough to count. For each of us who sees them will also form the concepts of three, four, and many. With the Moon, God established the number of days in a month, and then months in relation to the year. Thanks due to these celestial events we have crops, the earth bears food for all living things, and the winds that blow and the rains that fall are not violent or without measure. If on the contrary anything turns out for the worse, we must not blame God, but humans, for not rightly managing their own lives (*Epinomis,* 978c-979b).

Humans should admit as evidence of the intelligence of the stars, the fact that they do the same things, because they are doing what was decided long time ago and do not change their decision back and forth and changing their orbits. This opinion is the exact opposite of what most people believe, that because they do the same things uniformly they do not possess soul (*Epinomis*, 982c, d). If they are divine, they should be one of these two things: either they are themselves gods, or we must suppose them to be likenesses of gods, something like images of them, made by the gods themselves (*Epinomis,* 983e-984a). The gods, therefore, are the fixed stars, the Moon, the Sun and the planets. *Epinomis* discusses what things a person is to learn about reverence towards the gods and how he is to learn them. Those things pertain to 'astronomy', and this is an answer no one would ever expect through unfamiliarity with the subject. People do not know that the true astronomer must be the wisest person. However, astronomer is not the one who practices astronomy the way Hesiod did and everyone else of that sort, by observing risings and settings of stars; astronomer is the one who has observed seven of the eight circuits, each of them completing its own orbit in a way no one can easily contemplate who is not endowed with an extraordinary nature (*Epinomis,* 990a, b).

### 3. Discussion

The disciplines closely related to numbers intended as 'abstract' entities (or beyond the perceived reality), that is arithmetic, geometry and astronomy, are those really important for the education in the ideal city, in particular for the philosopher-rulers and Guardians of the Law. Plato is not worried by the fact that those disciplines would appear of no 'practical' utility. Even the calculations applied to practical problems in the various arts must be learned, but they are of secondary importance with respect to the disciplines that are required for a 'gentleman', who shall deal with the laws and the government of the city.

This distinction between 'practical' and 'theoretical' disciplines appears arguable in *Epinomis*, and a certain disesteem for farming, which in Plato's dialogues was maybe suggested by its low ranking among the human activities, becomes in a certain sense annoying in his school. The sentence: "it is clearly not by art but by a natural capacity we have from God that we have all put our hands to working the earth" could be questionable. Maybe a farmer would not have been in agreement, since, as any other art and craft, also farming must be learned; one can give a look at the Latin treatises, e.g. *De Re Rustica* by Columella (4 – 70 CE), on that subject. More generally, in that epoch there was the belief that, in the evolution of humankind, farming was preceded by a different lifestyle. During the golden age, the earth supplied directly fruits and food, and it was not needed working it. Therefore, the ability to work the earth should have been acquired, and could not be innate. In the myth of the origin of humankind in the *Republic* (III, 414d – 415a), the men born directly from the ground with their specific abilities and tools; there was the class of the rulers, that of auxiliaries, and the less valued one of craftsmen and farmers. That is, the art of working the earth should be innate at most only in some persons.

According to *Epinomis*, it is thanks to the celestial events that we have crops and the earth bears food for all living things; but if on the contrary anything turns out for the worse, we must blame humans, for not rightly managing their own lives. Once more, presumably a farmer would not have been in agreement; but perhaps he could not reply, since probably he was illiterate. According to Plato (*Republic*, II, 379c), since a god is good, he is not – as most people claim – the cause of everything that happens to human beings but of only a few things, for good things are fewer than bad ones in our lives. He alone is responsible for the good things, but we must find some other cause for the bad ones. What was the cause of bad things? Plant and animal diseases, prolonged drought, locust invasions and similar disasters probably were not up to the poor farmer for not rightly managing his own life. They might derive from celestial events; see for example the "morbo caeli", infection of heavens or sickened sky, mentioned by Virgil (70 – 19 BC) in *Georgica* (III, 478), and the resulting, impressive devastation. Virgil wrote: "Not so thick with driving gales sweeps a whirlwind from the sea, as swarms scourges swarm among cattle. Not single victims do diseases seize, but a whole summer's fold in one stroke, the flock and the hope of the flock, the whole race, root and branch. Of this may one be witness, should he see—even now, so long after—the towering Alps and the forts on the Noric hills, and the fields of Illyrian Timavus with the shepherds' realm derelict, and their glades far and wide untenanted" (*Georgica*, III, 470-477).

In Plato's *Symposium*, one of the participants said: "But when the sort of Love that is crude and impulsive controls the seasons, he brings death and destruction. He spreads the plague and many other diseases among plants and animals; he causes frost and hail and blights. All these are the effects of the immodest and disordered species of Love on the movements of the stars and the seasons of the year, that is, on the objects studied by the science called astronomy" (*Symposium*, 188a, b). Plato affirmed clearly in the *Laws* that "the all-controlling agent in human affairs is God, assisted by the secondary influences of 'chance' and 'opportunity'. A less uncompromising way of putting it is to acknowledge that there must be a third factor, namely 'skill', to back up the other two" (*Laws,* IV, 709b, c). In the antiquity, a farmer could make only sacrifices to propitiate the gods and escape the disasters. Horace in *Odes* (III, 23) described a moving scene of a sacrifice performed by a simple-minded country girl, wishing her that her vine be fertile and not feel the wind which brings disease from Africa, nor the crop know the blight of mildew nor the lovely suckling beasts know a time of danger. The sacrifices for good harvest, however, could be bloody or even tragically human, as described in detail for instance by Frazer (1922; Ch. XLVII).

The 'distinction' among human activities, i.e. to exalt one activity and point out the defects of the others, is a very old one. The '*satire of the trades*' is a work of didactic literature of ancient Egypt (2$^{nd}$ millennium BC) showing the advantages of the profession of scribe while putting in negative light the other arts. Even in the Bible, in the *Book of Sirach* or *Ecclesiasticus* of 2$^{nd}$ century BC it is written (in a way that at the beginning it reminds of Plato's thought; *Republic*, II, 374d, e): "A scribe's wisdom is in the opportunity for leisure, and he who does less business, it is he who will become wise. How shall he who takes hold of a plow and boasts in the shaft of a goad become wise, when he drives cattle and is engaged in their tasks and his talk is about the offspring of bulls?" Similarly, the other artisans, smiths and potters cannot become wise. "All of these relied on their hands, and each is skilled in his work. Without them a city will not be inhabited, and they will neither sojourn nor walk about, but they will not be sought for a council of the people" (*Sirach*, 38, 24-34).

We think that, in general, there was a separation between the culture of scholars and the practical activities. In the following, after the case of Hesiod, who maybe had at least a taste of the hard work of the farmer before writing a beautiful poem on that topic, we will report other examples taken from the Greek and Latin literature. They will show in particular some astronomical inconsistencies in the writings of scholars who wished even to teach the 'practical' astronomy to the illiterate farmers.

## 4. Hesiod and Aratus

Hesiod's *Works and Days* contains a good example of a farming calendar of $8^{th}$ century BC. Without such a calendar, it would not have been possible to get enough (and good) crops to sustain the farmer and his family. It is easy to compare the few astronomical events (fixed star phases) used by Hesiod with those estimated by means of computing programs that take into account the dependence on the latitude of the locality (in Bœotia), the precession and the visibility conditions (e.g. Schaefer, 1985; Antonello, 2013a). The dates are consistent with what is known about the farming techniques in past millennia. A reference point for the time reckoning was the winter solstice. That is shown for instance by the sentence "When Zeus has finished sixty wintry days after the solstice" (*Works and Days*, l. 564), which indicated the beginning of the works after the winter season. That could be a reminiscence of what had been done presumably for millennia by the pre-literate civilizations, for example with megaliths oriented towards the rising or setting of the Sun at the winter or summer solstice, using them as a reference for starting the yearly time reckoning.

It seems that this practical astronomical culture became a bit confused when other scholars wrote poems and treatises on this subject. That is, the link between reasonably precise astronomy and poetry that can be noted in Hesiod was no more so evident. For example, it is known that the poem *Phenomena* of Aratus (about 315 - 240 BC) contains astronomical errors that were noted by Hipparchus (*Commentary on the Phaenomena of Eudoxus and Aratus*), and, in particular, it is known that it does not describe the visible sky at his or Eudoxus' time, but the sky of several centuries earlier[1]. However, as noted by Cicero in *De oratore* (I, XVI, 69), it was agreed in learned circles that Aratus, who knew no astronomy, had sung of the heavenly spaces and the stars in verse of consummate finish and excellence. Hence, the lack of precision of Aratus' astronomy apparently did not matter; the important thing was the quality of his poetry.

In the following Sections we will report some examples of the use of heliacal and achronycal risings and settings of stars, without entering into technical details. We recall that Neugebauer (1975) described such fixed star phases (pp. 1090-1091), mentioned the extensive literature on the parapegmata (meteorological calendars) or weather prognostications related with those phases (pp. 587-589), and discussed the theoretical work by Autolycus (about 360 – 290 BC) on this topic (pp. 760-763). Tannery (1886), in his work on this astronomer, noted that the risings and settings of stars had been one of the main interests of the early Hellenic astronomy, given the relation with the farming activities and the supposed dependence of the weather, and it is possible to conclude that ancient Greeks had a practical sidereal year, independently of their civil lunisolar year (see however Neugebauer 1975, pp. 1075-1076). Therefore, it seems that the confusion began when the scholars attempted to fit the fixed star phases, based implicitly on a sidereal year, into the civil calendar.

---

[1] This point was already discussed in $18^{th}$ century; see Lalande (1771, Vol. 2 pp. 339-340). Whiston and Fréret criticized the *Chronology* of Newton, in particular his fancy hypothesis about the so-called sphere of Chiron, that corresponds to the sky described by Eudoxus and Aratus. The authors estimated the epoch of the sphere about 1353 BC, while Newton's result had been 936 BC. An interesting indication, based on the comparison of observations of stars performed at different times, is that obtained by the astronomer Maraldi (1735) in 1733, who estimated an epoch of about 12 centuries BC, which is comparable with the result obtained recently by Schaefer (2004; 1130 ± 80 BC) with a thorough statistical analysis.

## 5. Pliny the Elder

In the *Naturalis Historia*, Pliny the Elder (23 - 79 CE) discussed the proper date for sowing the crops, that was in a large degree connected with astronomy, and he set forth the views that had been written in reference to the subject (XVIII, LVI, 201). Hesiod gave one date for sowing, the achronycal setting of the Pleiades, and still at Pliny's time that was the custom of sowing; we estimate that the date of the setting in autumn of the Pleiades was changed by about 11 days, due to the effect of the astronomical precession. There were, however, very different opinions about the corresponding right time of sowing, depending also, for example, on the different localities, soils and climate. According to Pliny, some writers did not pay attention to Nature, while others did pay too much, and consequently their elaborate theorizing was all in the dark, as the issue lay "between countrymen and literary, not merely astronomical, pundits!" (XVIII, LVI, 205-206). In other words, since the conclusions about the dates were unclear, Pliny pointed out the need to pay some attention, because the farmers were not just ignorant of astronomy, but also of learning. He wrote that it was "an arduous attempt and a vast aspiration - to succeed in introducing the divine science of the heavens to the ignorance of the rustic", but it should have been attempted, owing to the vast benefit it conferred on life. Nevertheless one had "first submit to contemplation the difficulties of astronomy"; even the experts themselves were conscious of those difficulties (XVIII, LVI, 205-206). It was "almost impossible to explain the system of the actual days of the year and that of the movement of the sun" (XVIII, LVII, 207). There had been great schools of astronomy that studied how to regulate the year in conformity with the Sun's revolution. Sosigenes himself, though more careful in research than the other writers, did not hesitate to introduce an element of doubt by correcting his own statements. Other authors published theories, although it was seldom that the opinions of any two of them agree (XVIII, LVII, 212-213).

Pliny pointed out a twofold difficulty: first of all, one had to ascertain whether or not the celestial phenomena were regulated by certain laws, and then one had to seek how to reconcile those laws with apparent facts. Then he remarked that one had to take into account the convexity of the earth, and the differences of situation in the localities upon the face of the globe. That difficulty had been considerably enhanced by authors through their observations having been taken in different regions, and because they actually published different and contradictory results of observations made in the same regions (XVIII, LVII, 210)[2]. Pliny did not mention the sky visibility in Africa, Spain, and the provinces of Gaul, since no one had published any observations made upon the stars in those countries. He remarked, however, that there would have been no difficulty in calculating them, since the rising of the heavenly bodies was the same for all parts within those parallels (XVIII, LVII, 217), that is, for localities with the same latitude. Anyway, Pliny recognized that the life led by the ancients was rude and illiterate, and still the observations they made were not less remarkable for ingenuity than were the theories of his times (XVIII, LXIX, 284).

Since the goal of Pliny was an attempt to give a general description, it is surprising he did not mention the precession. Discrepancies in the dates originated also from the effects of precession, when putting together (or when comparing) observations taken at intervals of many centuries. According to Evans (1998, p. 262) "precession appears never to have become very widely known in antiquity. It is never alluded to by Geminus, Cleomedes, Theon of Smyrna, Manilius, Pliny, Censorinus, Achilles, Chalcidius, Macrobius, or Martianus Capella. Pliny had an unquenchable thirst for astonishing facts, and he professed the highest admiration for Hipparchus. Had Pliny heard of precession he certainly would have mentioned it". According to Evans, the only ancient writers who mentioned the precession besides Ptolemy were Proclus, who denied its existence, and Theon of Alexandria[3]. However, we note that Columella mentioned Hipparchus and something related to the precession: the time will come when the poles will change position[4].

---

[2] Tannery (1886) pointed out that: "au temps d'Eudoxe, les observations de phases, faites par différents observateurs à différentes latitudes, sur l'horizon apparent plutôt que sur l'horizon astronomique, devaient former un chaos à peine utilisable". According to Tannery, the theory of Autolycus had been a necessary step "avant d'arriver aux théories plus perfectionnées, qui d'ailleurs laissent, même encore aujourd'hui, assez à désirer" (p. 247).

In his farming calendar, Pliny quoted the astronomical indications of several Greek and Latin authors, such as Columella, whose book dedicated to farming contains a very detailed calendar (*De Re Rustica*, book XI, *Dies caelestes*). It was an uncritical collection of risings and settings of stars and of other astronomical features, and probably it was not based on observations made by the Author. In the past, scholars have tried several times to solve the inconsistencies of that calendar, proposing also specific interpretations (see e.g. Le Bœuffle 1964; 1993)[5].

As a conclusion, one may note that the ancient writers created some confusion, when they tried to give a coherent description of the calendars, based on the risings and settings of stars, a description that was actually intended to benefit the farmers.

**6. Other examples**

A story told by Ausonius may put some other light on the problem. Ausonius (310 - 395 CE) was a Gallo-Roman, Latin poet, teacher of rhetoric in Bordeaux (Burdigala), tutor of the emperor Gratianus, and consul. In one of the letters addressed to his pupil Paulinus (Saint Paulinus of Nola) he mentioned the bailiff (agent or administrator) of his estate in Gaul. The bailiff was from Greece. This person, often sorely disappointed by the light harvests, "came to hate the name of bailiff; and after sowing late or much too early through ignorance of the stars, made accusation against the powers above, carping at heaven and shifting the blame from himself. No diligent husbandman, no experienced ploughman, a spender rather than a getter, abusing the land as treacherous and unfruitful, he preferred to do businesses as a dealer in any-sale market" (Ausonius, *L. XVIII Epistularum*, XXVI). It may be possible that the bailiff had some knowledge of the farming techniques adopted in Greece, in that climate, and of the stars seen there, but in Gaul the situation should have been rather different in every respect. Presumably, similar difficulties should have occurred several times in the Roman Empire, when it was easy to move from one region (e.g. Syria) to a very far one (e.g. Britannia).

The *Geoponica* (or *Geoponika*) is an agricultural encyclopedia, a collection of agricultural lore, compiled in Constantinople for the Byzantine emperor Constantine VII Porphyrogenitus; the text in its present form dates from the mid-10th century. The sources are Hellenistic, Latin (such as Pliny the Elder), Roman-Greek, Carthaginian. It was very popular, and was translated in other languages of Near East. It includes many advices concerning cultivation and useful prognostics, based on the sky and the natural phenomena. Being an uncritical collection, one should expect inconsistencies as regards the 'practical' astronomy. "Since it is necessary that husbandmen should know the rising and setting of the apparent stars, I have written concerning them; so that persons wholly illiterate may, from memory, know the periods of their rising and setting" (*Geoponika*, I, *IX*). After that demanding declaration, the Author gave a list of dates for some bright stars (based on texts of Quintilii brothers; 2nd century CE). We have made some simulations for localities with latitudes between 32° and 44°, and for epochs between 500 BC and 1000 CE, as regards Pleiades, Arcturus, Regulus and Sirius. We have used a modified version of a program by Schaefer (1985; see Antonello 2013a, for some details). One should take however into account some possible uncertainties, such as the calendrical true date of the equinox when the observations were performed, and the possible presence of writing errors in the sources. In other words, the comparison must be intended only as qualitative. We assumed the spring equinox on 25th March, ("the vernal equinox is about the eighth of the calends of April"; *Geoponika*, I, *I*), even though it is not consistent with the corresponding astronomical event in the 10th century CE. The equinox was actually about on 16th March[6]; we show the results between square brackets for this case.

---

[3] See Neugebauer (1950; 1975 pp. 631-634).
[4] "Multos enim tam memorabiles auctores conperi persuasum habere longo aevi situ qualitatem caeli statumque mutari eorumque consultissimum astrologiae professorem Hipparchum prodidisse tempus fore, quo cardines mundi loco moverentur" (*De Re Rustica*, I, 1, 4).
[5] Useless attempts to solve the inconsistencies of the calendars of Varro (in *De Re Rustica*), Pliny and Columella were already performed in the past centuries as recalled by Lalande quoted by Saboureux (1783, pp. 95-99). The French author wrote that at the end he gave up reconciling the Latin authors.

According to *Geoponika* (I, *IX*) the heliacal setting of Pleiades was on the 1st of April (calends). The results of the program suggest that the date of such a setting of Pleiades is not very sensitive to the adopted latitude. In 10th century CE the setting was on 25th [16th] of April, in 5th century CE on 18th of April, and in 5th century BC on 4th of April. According to *Geoponika* (I, *IX*), the heliacal rising of Pleiades was on 7th of May or on 23rd of April; we find 29th [20th] of May for the 10th century CE, and a date earlier than the beginning of May only before the 5th century BC. Therefore, the rising and setting of Pleiades should had been actually observed several centuries BC. Analogously, according to *Geoponika,* the (heliacal) rising of Arcturus was on 15th of September, while we find this date some centuries BC. The estimated date of the heliacal rising of Regulus is 23rd [19th] of August in 10th century CE, and it is earlier than 7th of August before the 5th BC, while according to *Geoponika* it was on 30th of July. The heliacal rising of Sirius (20th of July) appears to be in agreement with observations performed in the first centuries CE.

There are other inconsistencies that would require a thorough discussion which is beyond the purpose of the present paper. On the whole, it seems that the dates for the stars are mostly compatible with a period of some centuries BC. In any case, one can see that the dates have no much to do with the risings and settings of stars in 10th century CE. Maybe, what suggested by *Geoponika* could have been a good memory exercise for the illiterate farmer, but completely unrelated with the farming activities.

## 7. Conclusion

We can conclude that, after Hesiod, the farming calendars with the risings and settings of the stars written by Greek and Latin authors appear to be merely an exercise of erudition, but they should have been of modest utility for the farmers. The separation between 'practical' and 'theoretical' astronomy discussed by us, however, suggests another, more general conclusion.

There is a common idea among scientists and humanists, that is, the human beings have had a particular relation with the heaven. Differently from that with the earth, the experience with the sky was possible only by means of sight. The effort to 'explain' what could be only seen, and not touched, should have been a powerful stimulus for the human thoughts and imagination[7]. For this reason we proposed to consider the sky-gazing as the 'mom' ('mum', 'mama', and not just the 'mother') of human knowledge (Antonello, 2013b). Sometimes it is possible to read sentences where mathematics, or philosophy, is defined as the mother of sciences, or the mother of knowledge. The sky-gazing should have occurred earlier than the development of mathematics and philosophy. A support to this view was taken from *Timaeus* and *Epinomis*, where the importance of celestial phenomena in this sense was clearly expressed, as we have seen in Section 2. In other words, sky-gazing could be considered the 'mom' of the various disciplines, not only of the scientific ones such as arithmetic, geometry, and astronomy itself, but also of those related to philosophy. We proposed to consider therefore cultural astronomy, intended as a sort of synonym of sky-gazing, as the place where the scientific and humanistic culture could really meet[8], given their common origin.

---

[6] According to the Latin authors, during the 1st century CE the spring equinox was about on 25th March. Modern estimates indicate 23rd – 22nd March for that epoch. Owing to the difference between Julian and mean tropical year (about 0.0078 day/year), the spring equinox was about on 21st – 20th March in the 4th century CE (First Council of Nicaea), on 16th – 15th March during the 10th century CE, and at the epoch of the Gregorian reform of the calendar (16th century) it was about on 11th – 10th March.

[7] It may be worth to mention also the history of religions. According to Ries (2007), right from the beginning the humankind was religious (*homo religiosus*), and, in agreement with Y. Coppens, at least since the time of *Homo habilis,* about two million years ago. That should have happened with the contemplation of the celestial vault. Given the anthropological importance of rites and symbols, in the case of a skeptical approach one should talk at least of *homo symbolicus*.

We think that the present study, aimed at pointing out the shortcomings of the culture separation within astronomy itself in the past, gives a further support to that proposal from a different point. Several politicians in Italy (and in other countries as well, I guess), given the persisting economic crisis, see the culture just as a luxury, i.e. it is a cost without benefit, and therefore it must be given a lower priority[9]. In Italy, some idiomatic expressions are frequently quoted when describing such a political attitude; we can summarize them with the sentence: culture doesn't feed people, where the verb 'to feed' must be intended literally. The question therefore could be: in the history of humankind, were the 'practical' astronomy and the 'theoretical' astronomy really essential for the human life? Surely, the 'practical' astronomy was of fundamental importance for farming, that is, the 'practical' astronomy was essential for supplying food and for feeding people. This is an obvious conclusion if we take into account what cultural astronomy and archaeoastronomy suggest us about the societies in the prehistory. 'Astronomy', however, is essential for feeding people even today. Farming depends on the seasons, and the seasons are an astronomical phenomenon. Farming depends on the climate, and the palaeoclimate studies show that the long-term climate changes are triggered by the orbital forcing, that is, the change of solar insolation related to that of Earth's orbital parameters. Given the complex behaviour of the climate system, those slow changes should be taken into account in any study aimed at predicting the future climate, and to understand the inescapable effects on the human life (e.g. Mohtadi et al. 2016). Only when putting together the results of the researches both in the natural and in the humanistic sciences (particularly those of the last few decades) it is possible to appreciate the fundamental importance of the 'astronomical' effects in the evolution of humankind and of her societies[10].

It would be wise do not subdivide 'astronomy' in sub-disciplines when talking to the public, and, in particular, to politicians; it would be better to consider its various branches (astrophysics, planetology, cosmology and so on), so to speak, just an internal affair. One should point out the specific importance of 'astronomy' in a broad sense, for the essential needs of the everyday life, and not only for cultural reasons.

---

[8] Snow (1961, p. 17) complained that there seemed to be no place for the meeting of the two cultures.

[9] Just as an example, as regards the basic research, a recent Nature editorial wrote that in democracies, politicians have to demonstrate to their electorates that they have not thrown away taxpayers' money on "self-indulgent frippery" (Nature, *Prove the worth of basic research*, 535, p. 465, 28 July 2016).

[10] The recent review by Mohtadi et al. (2016) is focussed on predicting monsoons, since they "have profound impacts on regions that are collectively home to more than 70 per cent of Earth's population"; reliable prediction of summer monsoons is critical to mitigating the often catastrophic consequences of rainfall anomalies, such as floods and droughts, famine and economic losses. In addition to the orbital forcing, Mohtadi et al. (2016) mention also the solar forcing, that is the possible effects of the change of solar activity; for a review of that specific topic, see Gray et al. (2010). Detailed discussions regarding the hominin evolution and that of human societies related to the effects of climate changes can be found for example in the reviews by Potts (2013) and Brooks (2006). We have tried to discuss briefly a possible 'astronomical perspective' in such a context (Antonello, 2013c).

Plato, *Republic*, trans. G.M.A. Grube, C.D.C. Reeve, in *Plato. Complete works*, J.M. Cooper (ed.), Hackett Publishing Company, Indianapolis, Indiana, 1997.

Plato, *Symposium*, trans. A. Nehamas, P. Woodruff, in *Plato. Complete works*, J.M. Cooper (ed.), Hackett Publishing Company, Indianapolis, Indiana, 1997.

Plato, *Timaeus*, trans. D.J. Zeyl, in *Plato. Complete works*, J.M. Cooper (ed.), Hackett Publishing Company, Indianapolis, Indiana, 1997.

Pliny the Elder, *Naturalis Historia*, in *Pliny. Natural History. Books 17-19*, trans. H. Rackham, Loeb Classical Library, Harvard University Press, Cambridge MA, 1950.

Potts R. (2013) *Hominin evolution in settings of strong environmental variability*, Quaternary Science Reviews, 73, pp. 1-13.

Ries J. (2007) *L'uomo religioso e la sua esperienza del sacro*, Opera Omnia Vol. III, Milano, Jaca Book; see https://en.wikipedia.org/wiki/Julien_Ries

Saboureux de la Bonneterie C.F. (1783) *Traduction d'anciens ouvrages latins relatifs à l'agriculture et à la médecine vétérinaire*, Tome Second, *L'économie rurale de Varron*, Paris, Barrois.

Schaefer B.E. (1985) *Predicting heliacal risings and settings*, Sky and telescope, 70, 261–263.

Schaefer B.E. (2004) *The latitude and epoch for the origin of the astronomical lore of Eudoxus*, Journal for the History of Astronomy, 35, 161 – 223.

*Sirach*, trans. B. G. Wright, rev. A.A. Di Lella, in *A New English Translation of the Septuagint*, Oxford University Press, New York, 2009.

Snow C.P. (1961) *The two cultures and the scientific revolution: The Rede Lecture 1959*, Cambridge University Press.

Tannery P. (1886) *Autolycos de Pitane*, in Mémoires Scientifiques, J.-L. Heiberg, H.-G. Zeuthen (eds.) Vol. II, Sciences exactes dans l'Antiquité 1883-1898, Paris, Gauthier-Villars (1912), 225-255.

Virgil, *Georgica*, in *Virgil. Eclogues, Georgics, Aeneid I-VI,* trans. H.R. Fairclough, rev. G.P. Goold, Loeb Classical Library, Harvard University Press, Cambridge MA, 1999.